\documentclass{appolb}
\usepackage{graphicx}
\usepackage{subcaption}
\usepackage{amsmath}
\usepackage{amssymb}
\usepackage{color}
\usepackage{slashed}
\usepackage{cite}
\usepackage{mciteplus}
\usepackage{braket}

\usepackage{hyperref} 

\begin{document}
\title{Recent developments in testable leptogenesis%
\thanks{Presented at Matter To The Deepest 2023.}%
}
\author{Yannis Georis
\address{Centre for Cosmology, Particle Physics and Phenomenology (CP3), 
Universit\'{e} catholique de Louvain, Chemin du Cyclotron 2, 
B-1348 Louvain-la-Neuve, Belgium}
\\[3mm]
}
\maketitle
\begin{abstract}
Low-scale leptogenesis is an attractive explanation for the observed baryon asymmetry of our universe that can be tested at a variety of laboratory experiments. In these proceedings, we review some recent advances in this field. In particular, we find that the viable parameter space is strongly enhanced, compared to the minimal case with two right-handed neutrinos, when a third generation is considered and explore the impact of such enhancement on the testability of the scenario. Finally, we also look at the impact of specific flavour and CP symmetries on said parameter space.
\end{abstract}
  
\section{Introduction}
Over the last decades, several hints of physics beyond the Standard Model (SM) have emerged. Among the most significant ones, the origins of neutrino oscillations as well as of the observed asymmetry between matter and antimatter, parametrised by the baryon-to-entropy ratio $Y_{B,\rm{obs}} \equiv \frac{n_B - n_{\bar{B}}}{s} \simeq 8.7 \cdot 10^{-11}$, cannot be explained within this framework. Extending the SM with right-handed neutrinos (RHN) provides a minimal solution to both of these problems. However, RHN masses and couplings are barely constrained from a theoretical viewpoint and their possible range spans several orders of magnitude. Motivated by the large number of experiments \cite{Abdullahi:2022jlv} currently searching for new physics at or below the TeV scale, we focus in these proceedings on such low-scale models. 

\section{The type-I seesaw mechanism}
At the renormalisable level, right-handed neutrinos $\nu_{Ri}$ can only interact with SM fields through the Lagrangian
\begin{eqnarray}
	\mathcal L \supset
	\mathrm{i} \, \overline{\nu_{Ri}} \, \slashed\partial \, \nu_{Ri}
	- \frac{1}{2}
	\overline{\nu^c_{Ri}}\, \left(M_{M}\right)_{ij}\, \nu_{Rj}
	- F_{\alpha i} \, \overline{l_{L\alpha}} \, \tilde{H} \, \nu_{Ri}
	+ {\rm h. c.} \, 
\label{eq:typeIseesawLagrangian}
\end{eqnarray}
where $M_M$ is the RHN Majorana mass matrix and $F$ is the Yukawa coupling matrix between the RHNs, the lepton doublet $l_L$ and the Higgs field $\tilde{H} = i\sigma_2 H$. After the electroweak symmetry breaking, the Higgs expectation value $\braket{H} = v_{\mathrm{ew}}$ generates a mixing $\theta=v_{\mathrm{ew}} F  \cdot M_M^{-1}$ between the theory's light ($\nu$) and heavy ($N$) mass eigenstates 
\begin{equation}
m_\nu \simeq -v_{\mathrm{ew}}^2 F  \cdot M_M^{-1} \cdot F^t, ~~~~~~~m_N \simeq M_M
\end{equation}
through the so-called \textit{type-I seesaw mechanism}. These new heavy fermions $N$ are commonly referred to as \textit{heavy neutrinos}. Given their gauge singlet nature, the number of RHN generations $n$ remains theoretically unconstrained. However, the experimental requirement to reproduce the two observed SM neutrino mass splittings forces $n$ to be equal or larger than $2$. Since the minimal $n=2$ case has already been thoroughly studied in the literature, see \textit{e.g.} \cite{Asaka:2005pn,Canetti:2012zc,Klaric:2021cpi}, we will mostly explore in these proceedings the phenomenology associated with the next-to-minimal scenario where $n=3$. The latter number of generations is also very well motivated from a theoretical perspective as it naturally appears in \textit{e.g.} theories with flavour symmetries or where RHNs have gauge interactions. To conclude this section, we point out that, while the number of free parameters in the type-I seesaw \eqref{eq:typeIseesawLagrangian} grows with $n$ as $7n-3$, experimental sensitivities are usually expressed, see \textit{e.g.} Fig. \ref{fig:radish}, as a function of the three combinations
\begin{equation}
U_\alpha^2 \equiv \sum_{i} |\theta_{\alpha i}|^2 =  v_{\mathrm{ew}}^2 \sum_{i} |\left(F\cdot M_M^{-1}\right)_{\alpha i}|^2
\end{equation}
and the heavy neutrino masses only as these are the primary parameters governing the production and decay of heavy neutrinos at laboratory experiments.

\section{Low-scale leptogenesis}

Beyond being able to explain the neutrino masses' origin, heavy neutrinos can also provide the necessary additional C and CP violation as well as deviation from thermal equilibrium to source the baryon asymmetry of our universe (BAU). While Fukugita and Yanagida's initial idea \cite{Fukugita:1986hr} relied on the decay of one very heavy RHN ($M_M \gtrsim 10^{9}$ GeV), the attention has shifted in recent years \cite{Abdullahi:2022jlv} towards testable low-scale models of leptogenesis where the heavy neutrino masses lie below the TeV scale. As already mentioned, such scenarios are particularly attractive \cite{Abdullahi:2022jlv} as they can be tested at a large variety of experiments, from collider searches to fixed target or charged lepton flavour violation  experiments, see also Fig.~\ref{fig:radish}.

At such low scales, there exist two primary mechanisms to produce the BAU. 1) \textit{Resonant leptogenesis} \cite{Pilaftsis:1997jf,Pilaftsis:2003gt} where a mass degeneracy between two heavy neutrinos resonantly enhance the asymmetry produced during their decay. 2) \textit{Leptogenesis from neutrino oscillations} \cite{Akhmedov:1998qx,Asaka:2005pn} where RHN oscillations during their freeze-in generates the BAU. While these two scenarios have long been thought as completely distinct, it was recently shown for the minimal $n=2$ case \cite{Klaric:2020phc,Klaric:2021cpi} that the regimes where these mechanisms are efficient do actually overlap. In these two scenarios, an accurate description of heavy neutrino evolution in the early universe can only be achieved by solving a set of \textit{quantum kinetic equations} consistently including all thermal effects. A complete quantitative understanding of low-scale leptogenesis from a theoretical viewpoint has been the subject of intensive effort over the last two decades, see \textit{e.g.} \cite{Dev:2017wwc,Garbrecht:2018mrp,Laine:2022pgk} for reviews. In the following, we have used the set of equations and rates provided by \cite{Klaric:2021cpi}.

\subsection{Leptogenesis with $3$ RHN generations}

\begin{figure}[!t]
\hspace{-1.75cm}
\begin{subfigure}{0.6\textwidth}
\centerline{\includegraphics[width=\linewidth]{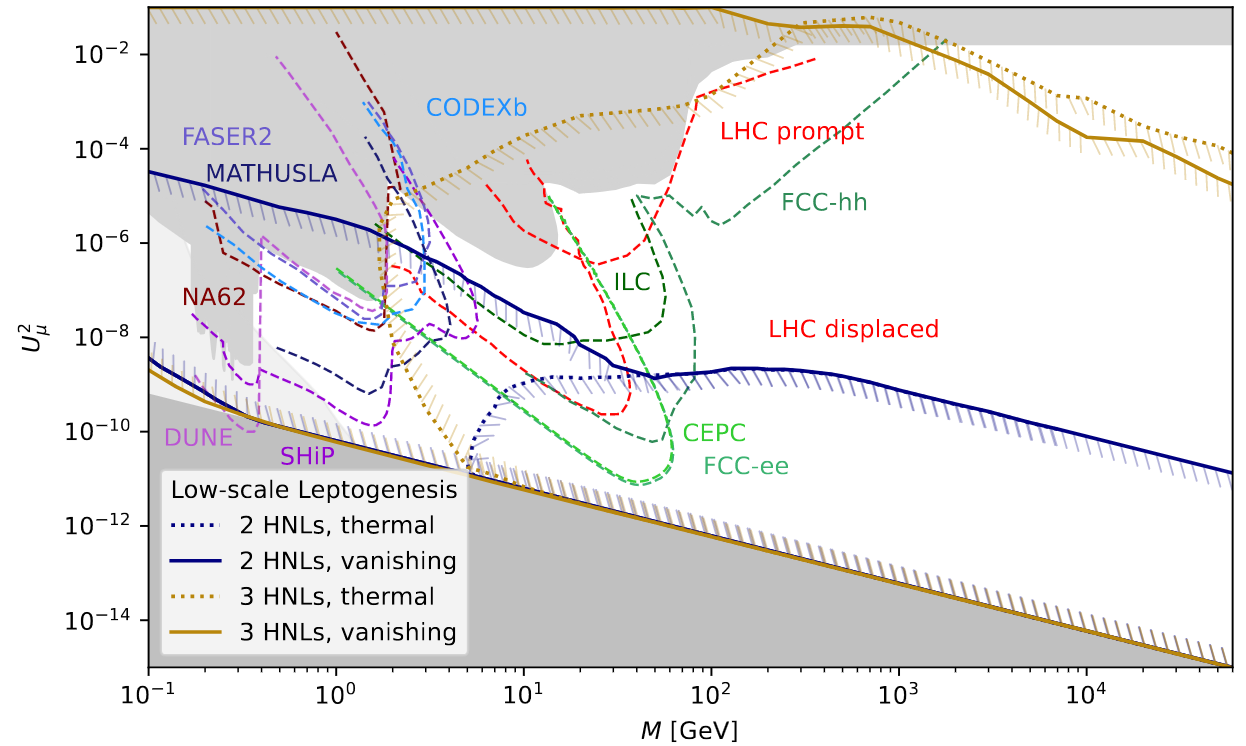}}
\end{subfigure}
\begin{subfigure}{0.565\textwidth}
\centerline{\includegraphics[width=\linewidth]{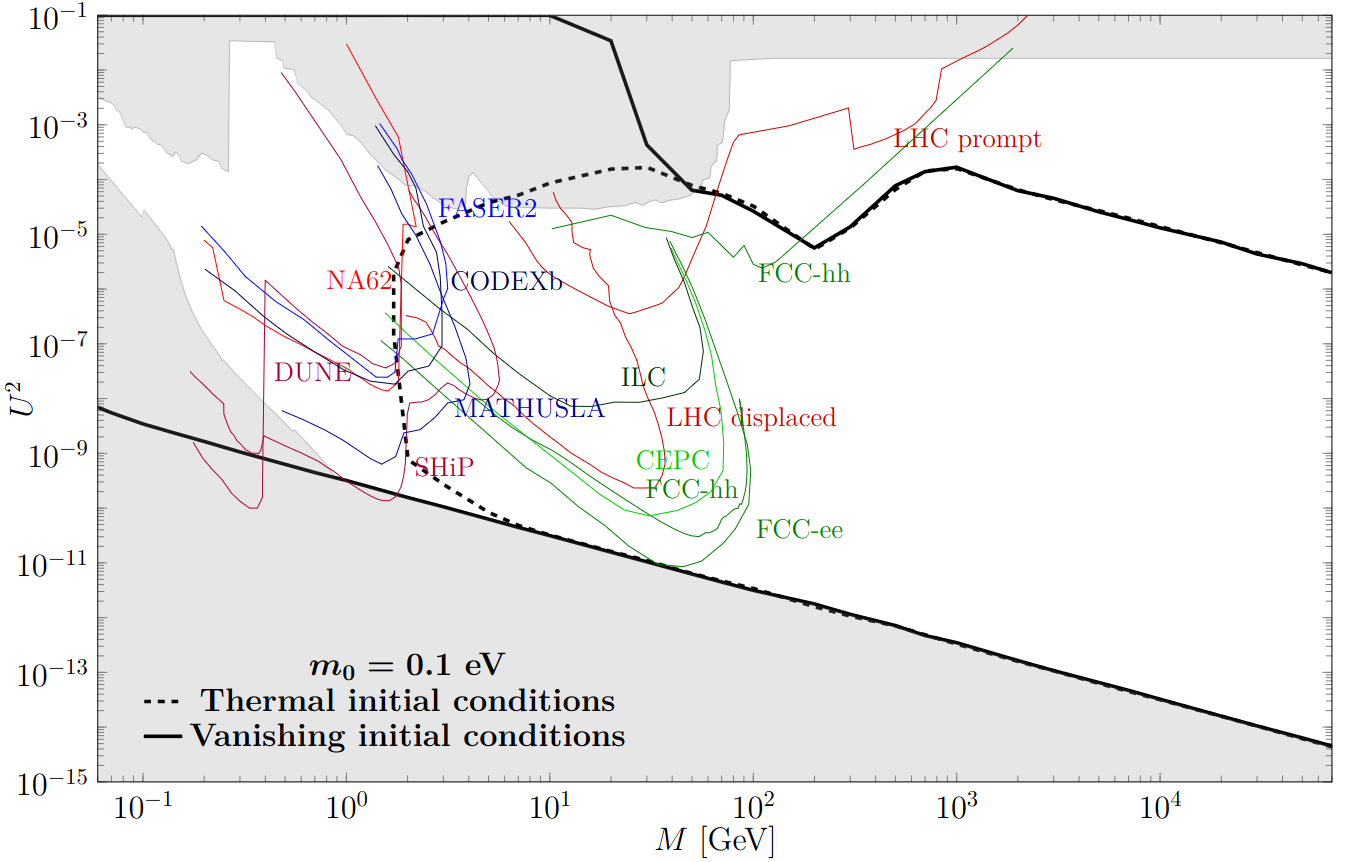}}
\end{subfigure}
\caption[]{The left figure displays the range for the heavy neutrino mixing with the muon neutrino $U_{\mu}^2$ compatible with both the observed neutrino masses and BAU as a function of the (average) heavy neutrino mass $M$ for both the $n=2$ (in blue) and the $n=3$ case (in orange) with $m_0=0$ eV. As initial condition, we either consider RHN to be absent (continuous lines) or in thermal equilibrium (dotted lines). The right plot shows the equivalent parameter space for $n=3$ and $m_0=0.1$ eV. Sensitivities of current experiments or proposals are displayed in color while the grey region is already excluded by a combination of constraints from direct experiments, Big Bang Nucleosynthesis and neutrino oscillation data. More details in Ref. \cite{Drewes:2021nqr,Abdullahi:2022jlv} from which we extracted both figures.}
\label{fig:radish}
\end{figure}

We first study the parameter space of the general type-I seesaw model with 3 RHNs that is consistent with leptogenesis which was for the first time mapped in \cite{Drewes:2021nqr}. As highlighted by Fig. \ref{fig:radish}, we observe that the viable parameter space is largely enhanced, \textit{i.e.} the possible range of heavy neutrino couplings to SM flavours is increased by several orders of magnitude compared to the $n=2$ scenario. Given this parameter was not yet measured, it is also interesting to look at the impact of the lightest SM neutrino mass $m_0$ on the parameter space. As displayed in the right panel of Fig.~\ref{fig:radish}, we observe that the viable parameter space only slightly decreases for $m_0 = 0.1$ eV. The observed dip around\footnote{We consider in our setup the 3 RHNs to be approximately degenerate $M_1 \approx M_2 \approx M_3 \approx M$ since the largest mixing angle $U^2$ will be reached for such scenarios.} $M=100$ GeV marks the transition between a regime where most of the asymmetry is produced during the RHN freeze-in to a regime where most of the BAU is generated during their freeze-out.

Such an enhancement of the viable parameter space bears promising consequences regarding the testability of the model. Indeed, in the most optimistic scenarios, one could not only discover but even potentially produce enough RHNs, more than $10^5$ at FCC-ee, to \textit{e.g.} measure with a percent-level precision \cite{Antusch:2017pkq} the flavour ratio $\frac{U_\alpha^2}{U^2}$. These additional information will allow us to perform consistency checks for the hypothesis that RHNs explain the origin of neutrino masses and the BAU. Such a large number of observed events would also open the ground for different types of tests such as \textit{e.g.} measuring the ratio between lepton number violating and conserving decays at colliders.

\begin{figure}[!t]
\hspace{-1.5cm}
\begin{subfigure}{0.6\textwidth}
\centerline{\includegraphics[width=\linewidth]{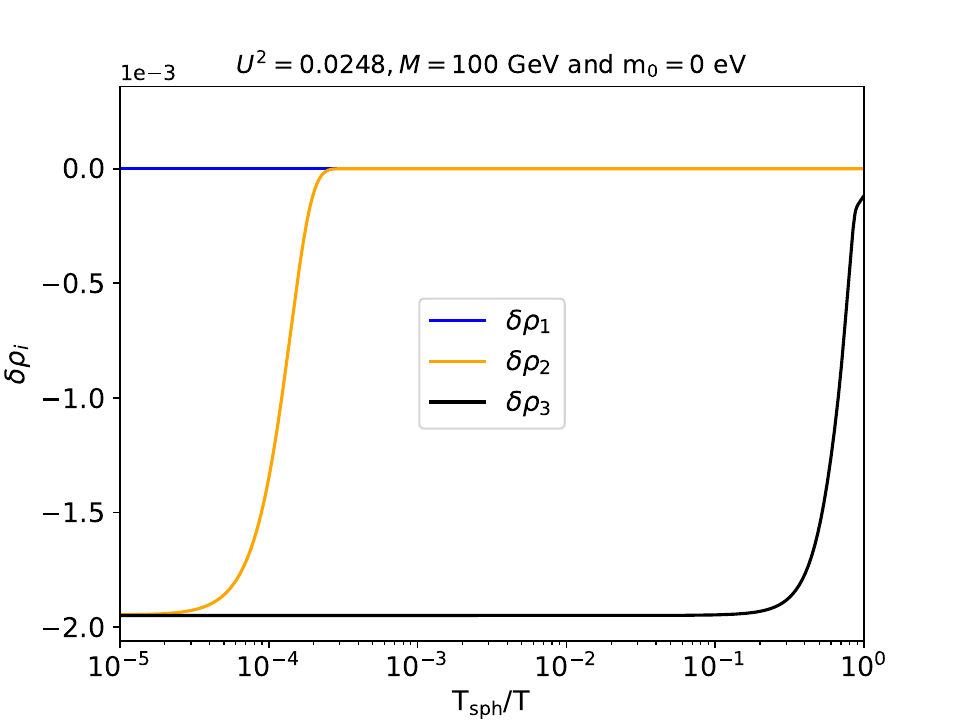}}
\end{subfigure}
\begin{subfigure}{0.6\textwidth}
\centerline{\includegraphics[width=\linewidth]{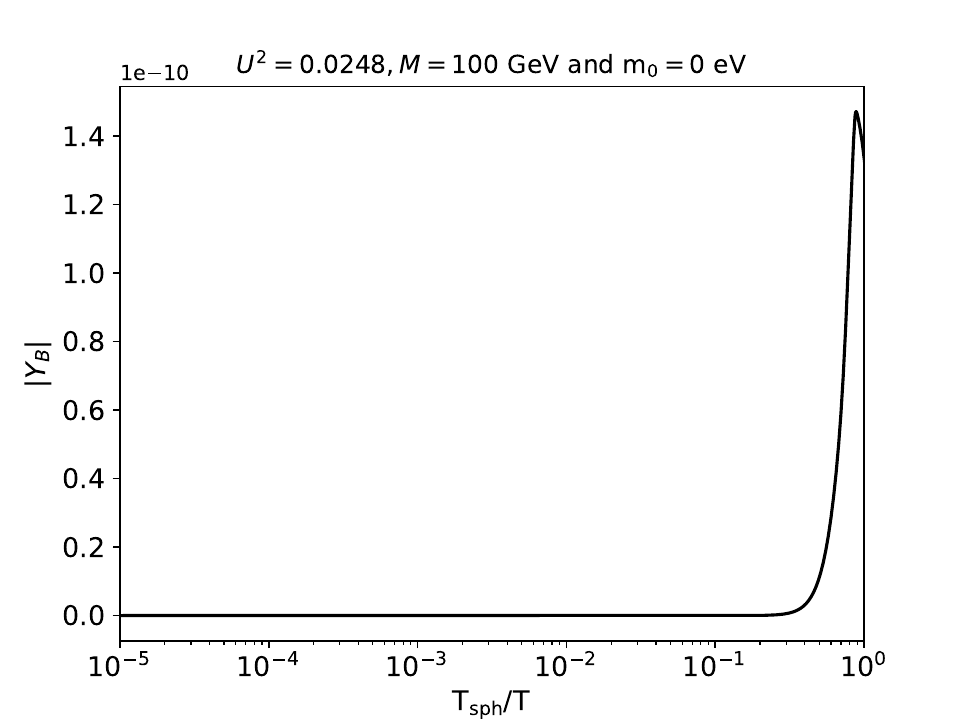}}
\end{subfigure}
\caption[]{Evolution of heavy neutrino abundances (\textit{left panel}) and the total BAU (\textit{right panel}) as a function of the temperature for a benchmark close to the upper bound displayed in Fig.~\ref{fig:radish}. Figure taken from Ref. \cite{Georis:2023skq}.}
\label{fig:LateBAUprodfreeze-in}
\end{figure}

The reason behind such an enhancement of the parameter space lies in the presence of a third RHN that is less constrained by the requirement to reproduce neutrino masses and can (almost) decouple. Hence, it can equilibrate at very late times, just before the freeze-out of the sphaleron. In that case, the asymmetry will also be produced right before sphaleron freeze-out and its washout will be limited. This feature is highlighted in Fig. \ref{fig:LateBAUprodfreeze-in}. While such a late time production might seem tuned, this scenario can actually be protected by an approximate B-L symmetry \cite{Shaposhnikov:2006nn}.

\subsection{Effects of flavour and CP symmetries}

Even though it can reproduce the observed neutrino masses and BAU, the type-I seesaw model \eqref{eq:typeIseesawLagrangian} does not however predict the various lepton mixing angles which have no dynamical origin in this scenario. In order to remediate to that, one can endow \eqref{eq:typeIseesawLagrangian} with flavour and CP symmetries. In particular, discrete symmetry groups have been shown to be extremely powerful to predict the values of the various phases and angles appearing in the PMNS matrix. We here consider the series of flavour symmetry groups $\Delta(6n^2)$, see \cite{Drewes:2022kap,Hagedorn:2014wha} for motivations and more details on the model. In general, additional scalars have to be introduced in order to dynamically generate the symmetry breaking. We nevertheless remain agnostic on the UV-completion of the scenario and assume that the temperature at which these new fields would have an influence is much higher than the relevant scale for leptogenesis. However, these flavour symmetries still leave an observable imprint on the low-scale dynamics by severely constraining the possible form for the Yukawa matrix $F$. In this scenario, leptogenesis is only sensitive to 6 or 7, depending on the exact setup, free parameters instead of 13 for the general type-I seesaw \eqref{eq:typeIseesawLagrangian}. Such reduction makes it possible to analytically understand the behaviour of the BAU as a function of all these parameters through the use of CP violating combinations, see \cite{Drewes:2022kap}.

\begin{figure}[!t]
\hspace{-1.25cm}
\begin{subfigure}{0.6\textwidth}
\centerline{\includegraphics[width=\linewidth]{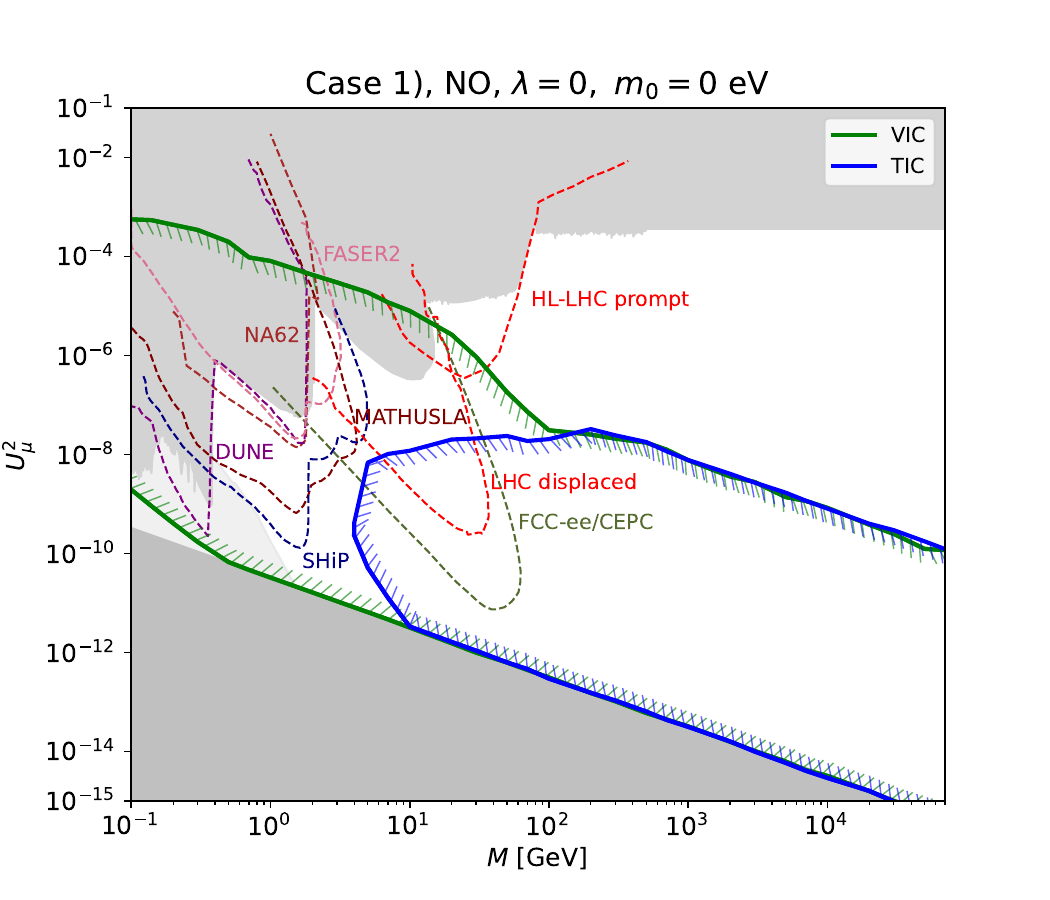}}
\end{subfigure}
\begin{subfigure}{0.6\textwidth}
\centerline{\includegraphics[width=\linewidth]{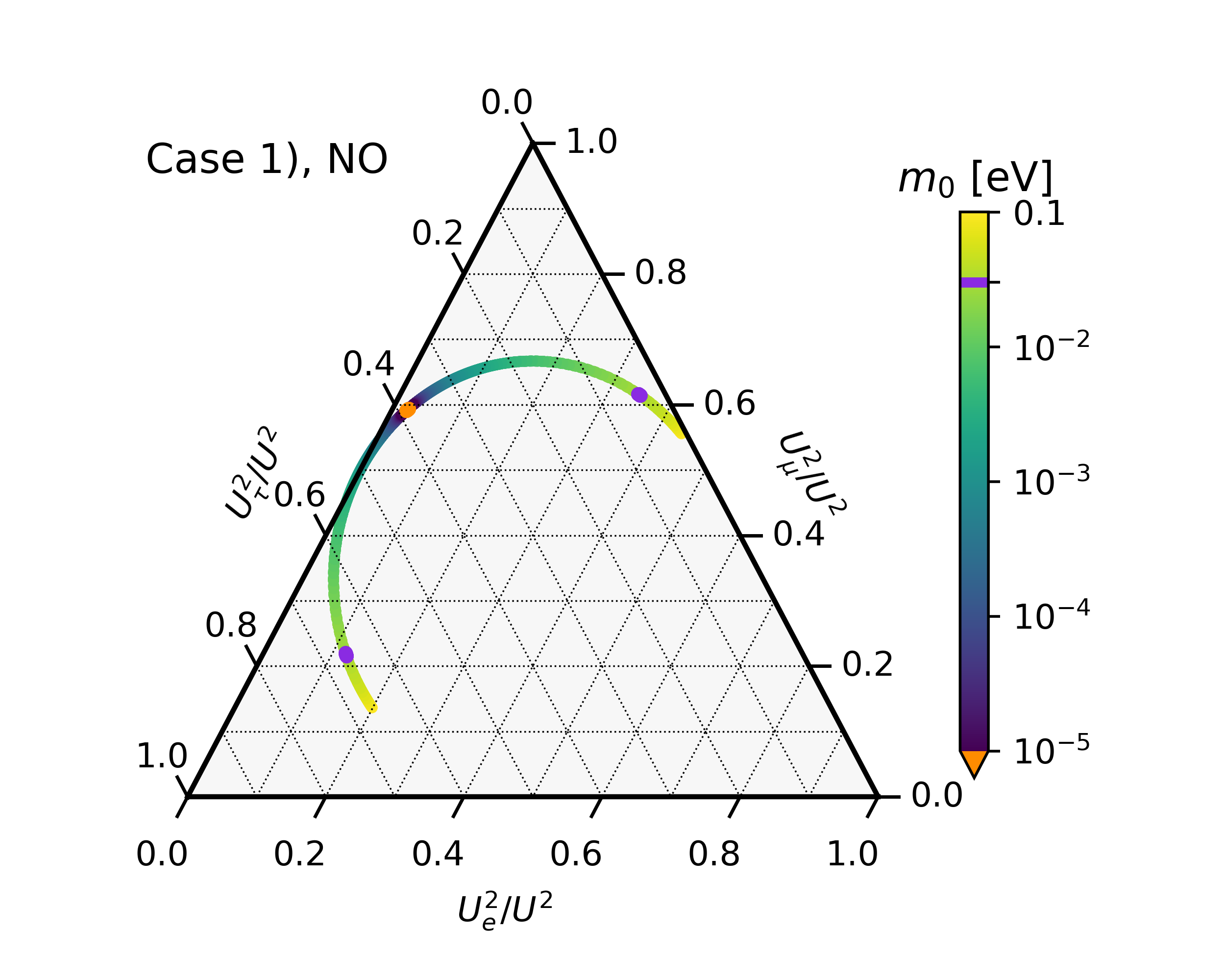}}
\end{subfigure}
\caption[]{The left plot describes the range of heavy neutrino mixing consistent with both leptogenesis and neutrino oscillation data for one specific scenario, labelled as Case 1) \cite{Drewes:2022kap}. The right plot displays the possible flavour ratio $\frac{U_\alpha^2}{U^2}$ for the same scenario for all possible choices of $m_0$. To appear in \cite{drewes:2024kkk}.}
\label{fig:Case1flavoursymmetry}
\end{figure}

Including flavour and CP symmetries, we observe that the parameter space is strongly reduced compared to the general type-I seesaw but remains testable by many experiments as can be seen from Fig.~\ref{fig:Case1flavoursymmetry}. In addition, flavour symmetries also drastically enhance the predictivity of the model. As an illustrative example, we show in the right plot of Fig.~\ref{fig:Case1flavoursymmetry} the range of branching ratio $U_\alpha^2/U^2$ compatible with neutrino oscillation data. We can observe that, once the lightest neutrino mass is fixed, at best only 2 choices are possible for these ratios which should be easy to rule out in case of detection. While not all cases are as predictive as the one displayed in Fig.~\ref{fig:Case1flavoursymmetry}, it remains generically valid that predictivity is largely improved. 

\section{Conclusion}

Extending the SM with RHN can explain the origin of neutrino masses and of the observed asymmetry between matter and antimatter in a minimal manner. In these proceedings, we reviewed recent studies of the viable parameter space for scenarios with 3 RHN generations. In the general type-I seesaw, we showed that the parameter space increases by several orders of magnitude, thereby largely enhancing testability prospects. We also showed that flavour symmetries, while reducing the parameter space, drastically enhance the predictivity of the model and could be more easily ruled out in case of detection.

\section*{Acknowledgments} 
The author would like to thank Marco Drewes, Claudia Hagedorn and Juraj Klari\'c for the long-term collaborations on the various projects mentioned in these proceedings and especially Claudia Hagedorn for many useful comments on the draft. He also acknowledges the support of the French Community of Belgium through the FRIA grant No. 1.E.063.22F.

\bibliographystyle{IEEEtran}
\bibliography{biblio}

\end{document}